\documentclass{article}
\usepackage{spconf,amsmath,graphicx}

\usepackage{amsfonts,amssymb,amsmath}
\usepackage{color}
\usepackage{algorithm}
\usepackage{algpseudocode}
\usepackage{cite}
\usepackage{url}

\title{Narrowband Channel Estimation for Hybrid Beamforming Millimeter Wave Communication Systems with One-bit Quantization}
\name{Junmo Sung, Jinseok Choi, and Brian L. Evans\thanks{This work is supported by gift funding from Huawei Technologies.}}
\address{Wireless Networking and Communications Group \\ The University of Texas at Austin, Austin, TX USA}
\begin{document}
%\ninept
%
\maketitle
\begin{abstract}
Millimeter wave (mmWave) spectrum has drawn attention due to its tremendous available bandwidth. The high propagation losses in the mmWave bands necessitate beamforming with a large number of antennas. Traditionally each antenna is paired with a high-speed analog-to-digital converter (ADC), which results in high power consumption. 
A hybrid beamforming architecture and one-bit resolution ADCs have been proposed to reduce power consumption. However, analog beamforming and one-bit quantization make channel estimation more challenging. 
In this paper, we propose a narrowband channel estimation algorithm for mmWave communication systems with one-bit ADCs and hybrid beamforming based on generalized approximate message passing (GAMP).
We show through simulation that 
1) GAMP variants with one-bit ADCs have better performance than do least-squares estimation methods without quantization, 
2) the proposed one-bit GAMP algorithm achieves the lowest estimation error among the GAMP variants,
and 3) exploiting more frames and RF chains enhances the channel estimation performance.
\end{abstract}

\section{Introduction} % (fold)
\label{sec:introduction}
As cellular communication technologies consider adopting millimeter wave (mmWave) bands in need of tremendous available spectrum, communication systems need a large number of antennas to compensate the high propagation losses in such frequency bands. A corresponding number of radio frequency (RF) chains including high-speed analog-to-digital converters (ADCs) are to be paired with the antennas in a traditional sense,
%employ a large number of antennas and a corresponding number of radio frequency (RF) chains to compensate a propagation path loss. They also need high-speed analog-to-digital converters (ADCs) to exploit the tremendous available bandwidth of the mmWave spectra. 
which inherently leads to high power consumption. A hybrid digital and analog beamforming architecture and low-resolution ADCs, therefore, were proposed in order to reduce power consumption primarily caused by those high-speed ADCs. 
%MmWave communication systems with the hybrid beamforming architecture still have room for improvement in power efficiency if they have a fairly large number of RF chains. It is to employ low-resolution ADCs, and an extreme case is ADCs with a single bit resolution. 

%One-bit ADCs had drawn attention for replacement to high-resolution ADCs in a massive multiple-input multiple-output (MIMO) communication system. 
Low-resolution ADCs have recently been combined with hybrid beamforming. Hybrid beamforming with low-resolution ADCs yields comparable achievable rates to those systems with high-resolution ADCs in the low and medium SNR regimes \cite{Mo2017}. The hybrid and full-digital beamforming architectures with low-resolution quantizers are shown to alternatively achieve better spectral and energy efficiency trade-off depending on characteristics of system components \cite{Abbas2016}. Spectral and energy efficiencies can be further enhanced by employing resolution adaptive ADCs with proper bit allocation algorithms \cite{Choi2017,Choi2017icassp}.

The combination of a hybrid beamforming architecture and low-resolution quantizers intended to reduce power consumption, however, makes channel estimation in such systems more challenging. 
%Channel estimation, however, becomes more challenging due to compression and distortion of received training signals caused by the hybrid beamforming and low-resolution quantization. 
Prior work regarding channel estimation in the context of mmWave can be grouped into a MIMO hybrid beamforming architecture with perfect quantizers \cite{Alkh2014,Park2016,Venu2017icassp,Alkh2015icassp} or a MIMO fully-digital architecture with low-resolution quantizers \cite{Mo2016,Mo2014,Rusu2015}. Very few publications concern a system equipped with hybrid beamforming and low-resolution ADCs \cite{Rodriguez2016,mo2018tsp}. In \cite{Rodriguez2016}, The modified expectation-maximization algorithm is shown to yield acceptable channel estimation errors with low-resolution ADCs. 
% In \cite{jsungICC2018}, orthogonal matching pursuit is proposed to use for wideband channel estimation, and four-bit ADCs are shown to achieve close performance to infinite bit ADCs. 
In \cite{mo2018tsp}, generalized approximate message passing (GAMP) is proposed to use for wideband channel estimation, and four-bit ADCs are shown to achieve close performance to infinite bit ADCs at medium SNR. 
The channel estimation algorithms in \cite{Rodriguez2016,mo2018tsp}, however, are not specially designed for one-bit ADCs. Brief comparison of the mentioned prior work is given in Table~\ref{tab:prior_work}.
\begin{figure*}[ht]
	\begin{minipage}[b]{1.0\linewidth}
		\centering
		\centerline{\includegraphics[width=17cm]{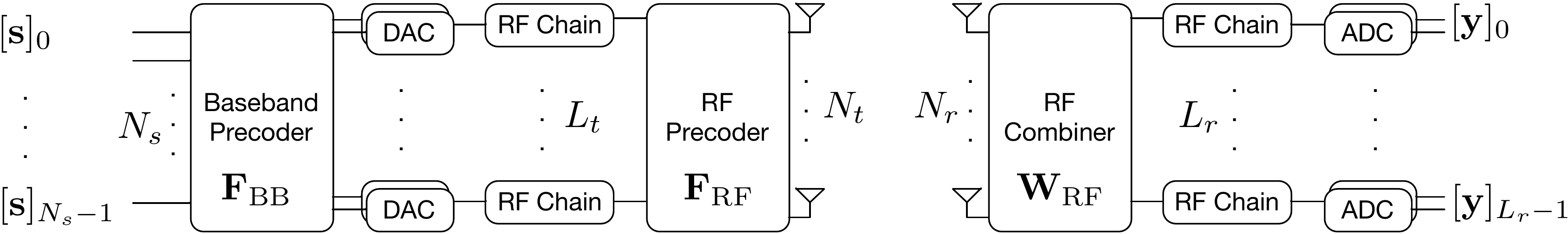}}
	\end{minipage}
	\caption{A block diagram of a MIMO hybrid beamforming communication system. A transmitter and a receiver in this system are equipped with $N_t$ and $N_r$ antennas, and $L_t$ and $L_r$ RF chains, respectively. }
	\label{fig:system_diagram}
\end{figure*}

In this paper, 
%we focus on one-bit quantizers unlike \cite{Rodriguez2016,jsungICC2018} where channel estimation algorithms are applicable to any quantization resolution. 
we propose a compressed sensing based channel estimation algorithm for a mmWave communication system equipped with one-bit ADCs and hybrid beamforming. GAMP and its variants have widely been used for channel estimation \cite{mo2018tsp,Mo2016,Mo2014,Garcia2016}. One variant of GAMP used in this paper is named one-bit GAMP and is specifically developed for measurements taken with one-bit quantizers \cite{Kamilov2012}. To make the algorithm better match a communication system model, we modify the algorithm to take the noise into account. Simulation results show that GAMP variants considered in this paper -- GAMP \cite{Rangan2011}, Expectation-Maximization Gaussian Mixture AMP (EM-GM-AMP) \cite{Vila2013}, and one-bit GAMP \cite{Kamilov2012} -- using one-bit ADCs perform better than does least-squares (LS) estimation without quantization. Among the variants, the modified one-bit GAMP achieves the lowest channel estimation error.
%Using the modified algorithm, we show that one-bit GAMP achieves better channel estimation performance than least-squares (LS) estimation without quantization
%the lowest channel estimation error among the widely used channel estimation algorithms that include least-squares (LS) estimation, GAMP and Expectation-Maximization Gaussian Mixture AMP (EM-GM-AMP). 
The results also show that estimation performance can be enhanced by exploiting more frames and RF chains.
%We use the modified algorithm for channel estimation and show that performance of the modified one-bit GAMP is superior to a traditional technique and GAMP variants, e.g. least-squares estimation, GAMP and Expectation-Maximization Gaussian Mixture AMP (EM-GM-AMP).
% section introduction (end)
\begin{table}[b!]
  \renewcommand{\arraystretch}{1}
  % \vspace{0.05in}
  \caption{Comparison of prior work}
  \label{tab:prior_work}
  \centering
  \begin{tabular}{l||l||l||l}
  \hline
  \bfseries Ref. & \bfseries Beamforming & \bfseries ADC Resolution & \bfseries Bandwidth\\
  \hline\hline
  \cite{Alkh2014} 		& Hybrid & Infinite & Narrowband \\
  \cite{Park2016} 		& Hybrid & Infinite & Narrowband \\
  \cite{Venu2017icassp} & Hybrid & Infinite & Wideband \\
  \cite{Alkh2015icassp} & Hybrid & Infinite & Narrowband \\
  
  \cite{Mo2016} 		& Digital & Low (1--9 bits) & Wideband\\
  \cite{Mo2014} 		& Digital & Low (1 bit) & Narrowband \\
  \cite{Rusu2015} 		& Digital & Low (1 bit) & Narrowband \\
  \cite{Rodriguez2016} 	& Hybrid  & Low (1--5 bits) & Narrowband \\
  \cite{mo2018tsp}	& Hybrid  & Low (1--4 bits) & Wideband \\
  \hline
  \end{tabular}
\end{table}

\section{System Model} % (fold)
\label{sec:system_model}
Consider a single-user MIMO hybrid beamforming mmWave system with $N_t$ transmit and $N_r$ receive antennas in the form of uniform linear arrays. The transmitter and receiver are equipped with $L_t$ ($\leq N_t$) and $L_r$ ($\leq N_r$) RF chains, respectively. The receiver employs quantizers that generate one-bit outputs. $N_s$ data streams are transmitted over narrowband MIMO channels where $ N_s \leq \min(L_t, L_r)$. The transceivers are assumed to have the same number of RF chains, i.e. $L_t=L_r$. The block diagram for the system is illustrated in Fig.~\ref{fig:system_diagram}. The baseband signal vector transmitted at $N_t$ antennas in the $m^{\text{th}}$ frame can be expressed as $\mathbf{x}_{m} = \mathbf{F}_{\text{RF}, m}\mathbf{F}_{\text{BB}, m}\mathbf{s}_m$ 
% \begin{align}
% 	\mathbf{x}_{m} &= \mathbf{F}_{\text{RF}, m}\mathbf{F}_{\text{BB}, m}\mathbf{s}_m,
% \end{align}
where $\mathbf{F}_{\text{RF}, m} \in \mathbb{C}^{N_t \times L_t}$ is the RF precoder, $\mathbf{F}_{\text{BB}, m} \in \mathbb{C}^{L_t \times N_s}$ is the baseband precoder, and $\mathbf{s}_m \in \mathbb{C}^{N_s \times 1}$ is the training symbol vector with the constraint $\mathbb{E}[\mathbf{s}_m\mathbf{s}_m^*]=\frac{1}{N_s}\mathbf{I}_{N_s}$. The RF precoder is assumed to be built using a network of analog phase shifters; therefore, all elements of $\mathbf{F}_{\text{RF},m}$ should have the identical norm of $\frac{1}{N_t}$. In order to control the transmit power, the baseband precoder has a constraint such that $\lVert \mathbf{F}_{\text{RF},m} \mathbf{F}_{\text{BB},m} \rVert_{F}^2=N_s$.
The quantized received baseband signal in the $m^{\text{th}}$ frame can be expressed as
\begin{align}
	\label{eq:received_signal}
	\mathbf{y}_m &= Q\left(\sqrt{\rho} \mathbf{W}_{\text{RF},m}^{\mathsf{H}} \mathbf{H} \mathbf{x}_m + \mathbf{W}_{\text{RF}, m}^{\mathsf{H}} \mathbf{n}_m\right),
\end{align}
where $(\cdot)^{\mathsf{H}}$ denotes the conjugate transpose, $\rho$ denotes the average received power, $\mathbf{W}_{\text{RF}, m} \in \mathbb{C}^{N_r \times L_r}$ is the RF combiner, $\mathbf{H} \in \mathbb{C}^{N_r \times N_t}$ is the flat-fading MIMO channel, and $\mathbf{n}_m \in \mathbb{C}^{N_r \times 1} \sim \mathcal{CN}(0, \sigma_n^2 \mathbf{I})$ is the noise vector. As with the RF precoder, the RF combiner has a constraint that all elements have the identical norm of $\frac{1}{N_r}$. Since quantization of the received signal in this system is performed with one-bit ADCs, the quantization operator $Q(\cdot)$ extracts signs of real and imaginary components of a complex argument. 
% The inverse quantization function, $Q^{-1}(\cdot)$, is defined as
% \begin{align}
% 	Q^{-1}(y) &=
% 	\begin{cases}
% 		[0, \infty) & \text{if } y=1   \\
% 		(-\infty, 0) & \text{if } y=-1.
% 	\end{cases} 
% \end{align}

\begin{figure}[t!]
	\begin{minipage}[b]{1.0\linewidth}
	  \centering
	  \centerline{\includegraphics[width=8cm]{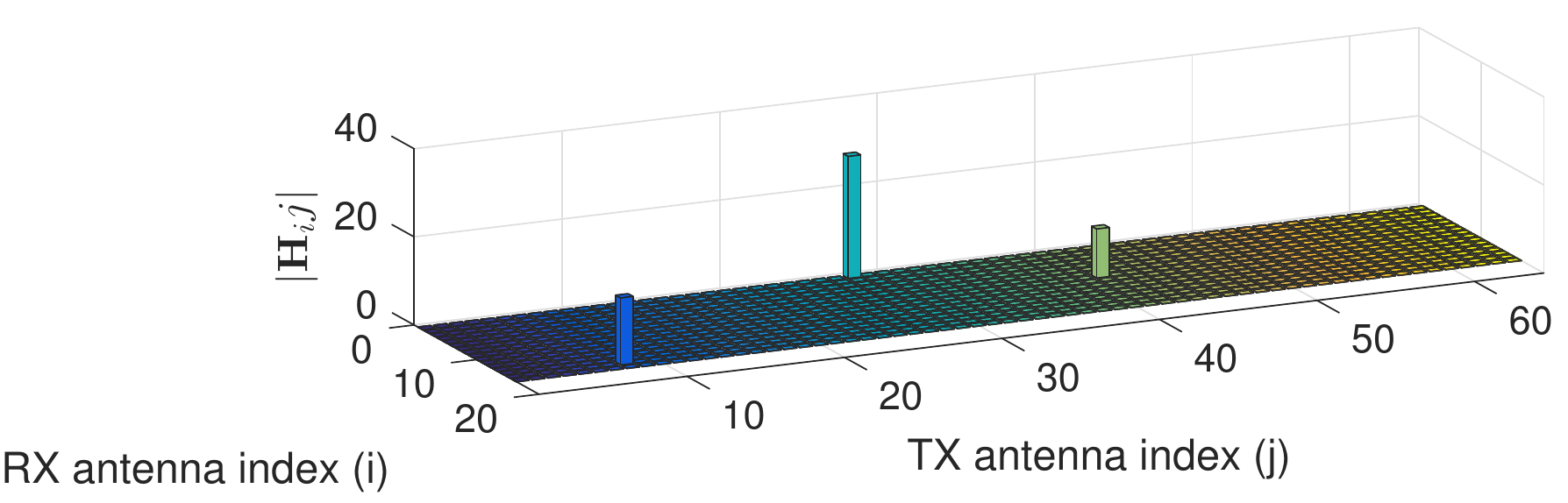}}
	%  \vspace{2.0cm}
	  \centerline{(a)}\medskip
	\end{minipage}
	\begin{minipage}[b]{1.0\linewidth}
	  \centering
	  \centerline{\includegraphics[width=8cm]{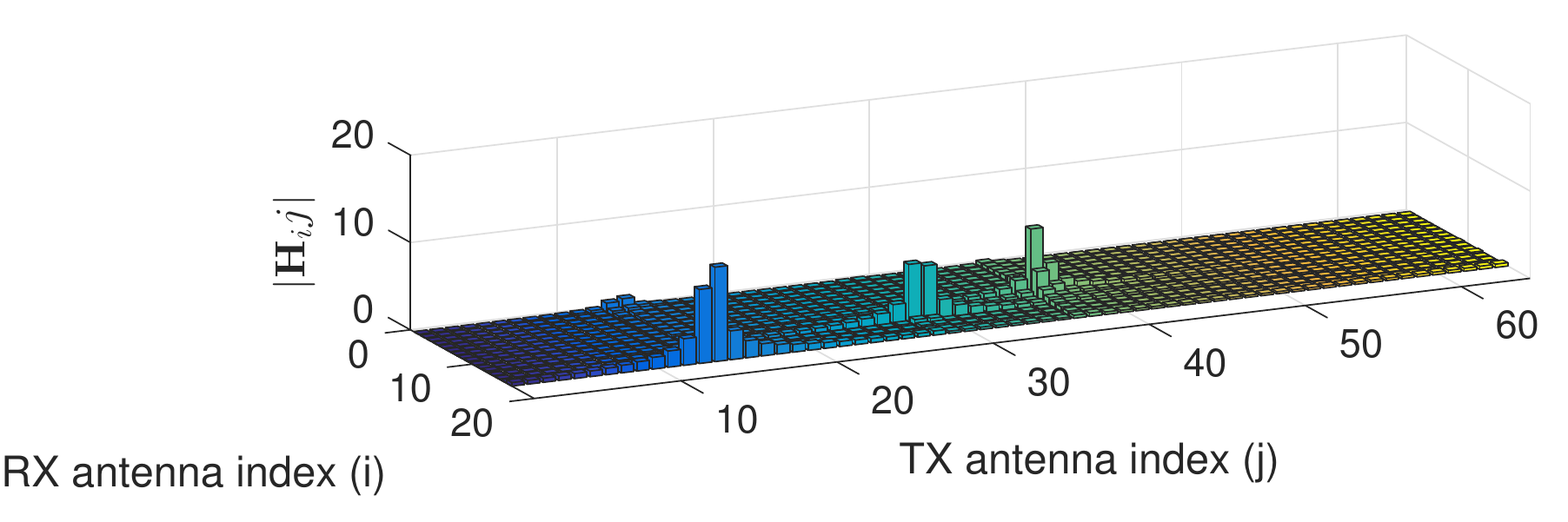}}
	%  \vspace{1.5cm}
	  \centerline{(b)}\medskip
	\end{minipage}
	% \hfill
	% \begin{minipage}[b]{0.48\linewidth}
	%   \centering
	%   \centerline{\includegraphics[width=4.0cm]{image4}}
	% %  \vspace{1.5cm}
	%   \centerline{(c) Result 4}\medskip
	% \end{minipage}
	%
	\vspace{-.3in}
	\caption{Magnitude of virtual channel matrix with $N_p=3$ in the angular domain. Spreads shown in (b) illustrate leakage into adjacent bins.}
	\label{fig:mag_Hv_ang_domain}
\end{figure}

Using a geometric channel model, the normalized channel matrix $\mathbf{H}$ can be constructed by summing up $N_p$ paths. The azimuth angles of departure and arrival (AoD and AoA) associated with the $l^{\text{th}}$ path are denoted as $\theta_{tl}$ and $\theta_{rl}$, respectively. Both $\theta_{tl}$ and $\theta_{rl}$ are uniform random variables distributed over  $[0, 2\pi)$. Therefore, $\mathbf{H}$ can be expressed as
\begin{align*} \
	%label{eq:channel}
	\mathbf{H} = \sqrt{\frac{N_r N_t}{N_p}} \sum_{l=0}^{N_p-1} \alpha_l \mathbf{a}_{\text{r}}(\theta_{rl}) \mathbf{a}_{\text{t}}^{\mathsf{H}}(\theta_{tl}),
\end{align*}
where $\alpha_l \sim \mathcal{CN}(0, \sigma_\alpha^2)$ is the complex channel gain of the $l^{\text{th}}$ path, $\mathbf{a}_{\text{t}}(\cdot) \in \mathbb{C}^{N_t \times 1}$ and $\mathbf{a}_{\text{r}}(\cdot) \in \mathbb{C}^{N_r \times 1}$ are the transmit and receive array response vectors at the given AoD and AoA, respectively. The channel matrix $\mathbf{H}$ is constrained to have $\mathbb{E}[ \left\lVert \mathbf{H} \right\lVert_F^2 ] = N_t N_r$ to maintain a constant channel power on average.
$\mathbf{H}$ can also be represented with the virtual channel representation as
\begin{align*} 
	\mathbf{H} = \mathbf{U}_{\text{r}} \mathbf{H}_{\text{v}} \mathbf{U}_{\text{t}}^{\mathsf{H}},
\end{align*}
where $\mathbf{U}_{\text{t}}$ and $\mathbf{U}_{\text{r}}$ denote the normalized $N_t$-point and $N_r$-point unitary Discrete Fourier Transform (DFT) matrices, and $\mathbf{H}_{\text{v}} \in \mathbb{C}^{N_r \times N_t}$ is the virtual channel matrix in the angular domain. As AoDs and AoAs are random in each channel realization, the array response vectors do not always align with columns of DFT matrices. This misalignment causes spectral leakage which degrades the performance of channel estimation algorithms. The leakage effect can be seen by comparing two subfigures in Fig.~\ref{fig:mag_Hv_ang_domain}. When AoDs and AoAs are perfectly aligned, or equivalently, antenna array response vectors can be expressed with the DFT matrix columns, $\mathbf{H}_{\text{v}}$ has exactly $N_p$ non-zero elements as seen in Fig.~\ref{fig:mag_Hv_ang_domain}(a). Otherwise, each beam leaks into adjacent bins, which results in spreads around them as shown in Fig.~\ref{fig:mag_Hv_ang_domain}(b). In Section~\ref{sec:numerical_results}, performance degradation due to the leakage effect is discussed.
% section system_model (end)

\section{Compressed Sensing Channel Estimation} % (fold)
\label{sec:compressed_sensing_channel_estimation}
For estimation of the sparse virtual channel, the $m^\text{th}$ received frame in \eqref{eq:received_signal} is reformulated into a vector form using $\text{vec}(\mathbf{A}\mathbf{B}\mathbf{C}) = (\mathbf{C}^{\mathsf{T}} \otimes \mathbf{A}) \text{vec}(\mathbf{B})$. Then $\mathbf{y}_m$ is rewritten as
\begin{align*} 
	\label{eq:vecForm_received_signal}
	\mathbf{y}_m &= Q \Big( 
	\sqrt{\rho} \left( \mathbf{s}_m^{\mathsf{T}} \mathbf{F}_{\text{BB}, m}^{\mathsf{T}} \mathbf{F}_{\text{RF}, m}^{\mathsf{T}} \otimes \mathbf{W}_{\text{RF}, m}^{\mathsf{H}} \right) \left( \mathbf{U}_{\text{t}}^{*} \otimes \mathbf{U}_{\text{r}} \right) \nonumber \\ 
	& \times \text{vec}(\mathbf{H}_{\text{v}}) + \tilde{\mathbf{n}}_m 
	\Big), 
\end{align*}
where $(\cdot)^{*}$ denotes the complex conjugation, $\otimes$ denotes the Kronecker product operator, and $\tilde{\mathbf{n}}_m$ denotes $\mathbf{W}_{\text{RF}, m}^{\mathsf{H}}\mathbf{n}_m$. For simple notation, we define
$\mathbf{\Phi} = [\mathbf{\Gamma}_0^{\mathsf{T}},\mathbf{\Gamma}_1^{\mathsf{T}},\cdots,\mathbf{\Gamma}_{M-1}^{\mathsf{T}}]^{\mathsf{T}}$, $\mathbf{\Psi} = \mathbf{U}_{\text{t}}^{*} \otimes \mathbf{U}_{\text{r}}$, and $\tilde{\mathbf{n}} = [\tilde{\mathbf{n}}_0^{\mathsf{T}},\tilde{\mathbf{n}}_1^{\mathsf{T}},\cdots,\tilde{\mathbf{n}}_{M-1}^{\mathsf{T}}]^{\mathsf{T}}$ 
% \begin{align} \label{eq:measurement_matrix}
% 	\mathbf{\Phi} &= 
% 	\begin{bmatrix}
% 		\mathbf{\Gamma}_0 \\ \mathbf{\Gamma}_1 \\ \vdots \\ \mathbf{\Gamma}_{M-1}
% 	\end{bmatrix},
% 	\mathbf{\Psi} = \mathbf{U}_{\text{t}}^{*} \otimes \mathbf{U}_{\text{r}}, \text{ and }
% 	\tilde{\mathbf{n}} = 
% 	\begin{bmatrix}
% 		\mathbf{\tilde{\mathbf{n}}}_0 \\ \mathbf{\tilde{\mathbf{n}}}_1 \\ \vdots \\ \mathbf{\tilde{\mathbf{n}}}_{M-1}
% 	\end{bmatrix},
% \end{align}
where $\mathbf{\Gamma}_m = \mathbf{s}_m^{\mathsf{T}} \mathbf{F}_{\text{BB}, m}^{\mathsf{T}} \mathbf{F}_{\text{RF}, m}^{\mathsf{T}} \otimes \mathbf{W}_{\text{RF}, m}^{\mathsf{H}}$. By stacking $M$ received frame vectors, $\mathbf{y}$ is obtained as
\begin{align*} 
	%\label{eq:simple_vecForm_received_signal}
	\mathbf{y} &= \left[\mathbf{y}_0^{\mathsf{T}},\mathbf{y}_1^{\mathsf{T}},\cdots,\mathbf{y}_{M-1}^{\mathsf{T}} \right]^{\mathsf{T}}
	% \begin{bmatrix}
	% 	\mathbf{y}_0 \\ \mathbf{y}_1 \\ \vdots \\ \mathbf{y}_{M-1}
	% \end{bmatrix}
	= Q \left( \sqrt{\rho} \mathbf{\Phi} \mathbf{\Psi} \mathbf{\widetilde{h}}_{\text{v}} + \tilde{\mathbf{n}} \right),
\end{align*}
where $\mathbf{\widetilde{h}}_{\text{v}}$ denotes $\text{vec}(\mathbf{H}_{\text{v}})$. We define the unquantized received vector $\mathbf{\widetilde{r}} = \mathbf{\widetilde{W}}\mathbf{\widetilde{h}}_{\text{v}} + \tilde{\mathbf{n}}$ 
% \begin{align}
% 	\mathbf{\widetilde{r}} = \mathbf{\widetilde{W}}\mathbf{\widetilde{h}}_{\text{v}} + \widetilde{\mathbf{n}}
% \end{align}
where $\mathbf{\widetilde{W}} = \sqrt{\rho} \mathbf{\Phi} \mathbf{\Psi}$. To separate in-phase and quadrature components, it is further reformulated as $\mathbf{r} = \mathbf{W}\mathbf{h}_{\text{v}} + \mathbf{n}$
% \begin{align}
% 	\mathbf{r} = \mathbf{W}\mathbf{h}_{\text{v}} + \mathbf{n}.
% \end{align}
where
\begin{align*}
	\mathbf{r} &= 
	\begin{bmatrix}
		\Re{\{\mathbf{\widetilde{r}}\}} \\ \Im{\{\mathbf{\widetilde{r}}\}} 
	\end{bmatrix}
	, \mathbf{W} =
	\begin{bmatrix}
 		\Re{\{\mathbf{\widetilde{W}}\}} & -\Im{\{\mathbf{\widetilde{W}}\}} \\
 		\Im{\{\mathbf{\widetilde{W}}\}} & \Re{\{\mathbf{\widetilde{W}}\}}
	\end{bmatrix}
	,\\
	\mathbf{h}_{\text{v}} &=
	\begin{bmatrix}
 		\Re{\{\mathbf{\widetilde{h}}_{\text{v}}\}} \\
 		\Im{\{\mathbf{\widetilde{h}}_{\text{v}}\}}
	\end{bmatrix} 
	, \text{and } \mathbf{n} = 
	\begin{bmatrix}
 		\Re{\{\tilde{\mathbf{n}}\}} \\
 		\Im{\{\tilde{\mathbf{n}}\}}
	\end{bmatrix}.
\end{align*}

\begin{algorithm}[t]
\caption{One-bit GAMP}\label{alg:1-bit_GAMP}
	\begin{algorithmic}[1]
	% \Procedure{One-bit GAMP}{}
	\State \textbf{Initialize:}
	\Statex \hspace*{\algorithmicindent}\parbox[t]{.8\linewidth}{$t=0$, $\hat{\mathbf{h}}_{\text{v}}^t = \mathbb{E}[\mathbf{h}_{\text{v}}]$, $\mathbf{v}^t_{\mathbf{h}_{\text{v}}} = \text{Var}[{\mathbf{h}_{\text{v}}}]$, $\hat{\mathbf{s}}^t=0$,}
	\For{$t=1, \cdots, T$}
		\Statex Measurement update:
			\State $\mathbf{v}^{t+1}_p = (\mathbf{W} \bullet \mathbf{W}) \mathbf{v}_{\mathbf{h}_{\text{v}}}^t$,
			\State $\hat{\mathbf{p}}^{t+1} = \mathbf{W}\mathbf{h}_{\text{v}}^t - \mathbf{v}_p^{t+1} \bullet \hat{\mathbf{s}}^t$,
			\For{all $i$}
				\State $\left[ \hat{\mathbf{s}}^{t+1} \right]_i = \frac{1}{[\mathbf{v}_p^{t+1}]_i + \sigma^2_{\tilde{n}}} ( \mathbb{E}\left[r|r \in Q^{-1}([\mathbf{y}]_i) \right] - $ 
				\Statex \hspace*{\algorithmicindent}\parbox[t]{.8\linewidth}{\hspace{55pt} $[\hat{\mathbf{p}}^{t+1}]_i )$,}
				\State $\left[ \mathbf{v}_s^{t+1} \right]_i = \frac{1}{[\mathbf{v}_p^{t+1}]_i + \sigma^2_{\tilde{n}}} \left( 1- \frac{\text{Var}\left[r|r \in Q^{-1}([\mathbf{y}]_i)\right]}{[\mathbf{v}_p^{t+1}]_i + \sigma^2_{\tilde{n}}} \right) $,
			\EndFor
		\Statex Estimation update:
			\State $\mathbf{v}_r^{t+1} = \left( (\mathbf{W} \bullet \mathbf{W})^{\mathsf{T}} \mathbf{v}_s^{t+1} \right)^{-1}$,
			\State $\hat{\mathbf{r}}^{t+1} = \hat{\mathbf{h}}_{\text{v}}^t + \mathbf{v}_r^{t+1} \bullet \left( \mathbf{W}^{\mathsf{T}} \hat{\mathbf{s}}^{t+1}\right)$,
			\For{all $i$}
				\State $\left[ \hat{\mathbf{h}}_{\text{v}}^{t+1} \right]_i = \mathbb{E}\left[ h_{\text{v}}|[\hat{\mathbf{r}}^{t+1}]_i \right]$,
				\State $\left[ \mathbf{v}_{\mathbf{h}_{\text{v}}}^{t+1} \right]_i = \text{Var}\left[h_{\text{v}}|[\hat{\mathbf{r}}^{t+1}]_i \right]$,
			\EndFor
	\EndFor
	% \EndProcedure
	\end{algorithmic}
\end{algorithm}

For sparse reconstruction of the virtual channel vector, we consider one-bit GAMP which is specifically developed for measurements taken with one-bit quantizers \cite{Kamilov2012}. We modify the proposed algorithm to take into consideration the additive noise and set the quantization threshold to zero. The modified algorithm is described in Algorithm~\ref{alg:1-bit_GAMP} where $\bullet$ denotes the element-wise product and $[\cdot]_i$ denotes the $i^{\text{th}}$ element in a vector. The expected value and the variance in lines 6 and 7 are with respect to $r \sim \mathcal{N}([\hat{\mathbf{p}}]_i, [\mathbf{v}_p^{t+1}]_i+\sigma^2_n)$. Those in lines 12 and 13 are with respect to $p_{h_{\text{v}}|[\hat{\mathbf{r}}^{t+1}]_i}(h_{\text{v}}|[\hat{\mathbf{r}}^{t+1}]_i)$ which is proportional to the product of $p_{h_{\text{v}}}(h_{\text{v}})$ and the Gaussian PDF with mean $[\hat{\mathbf{r}}^{t+1}]_i$ and variance $[\mathbf{v}_r^{t+1}]_i$.
% section compressed_sensing_channel_estimation (end)
\begin{figure}[t]
	\begin{minipage}[b]{1.0\linewidth}
	  \centering
	  \centerline{\includegraphics[width=7.4cm]{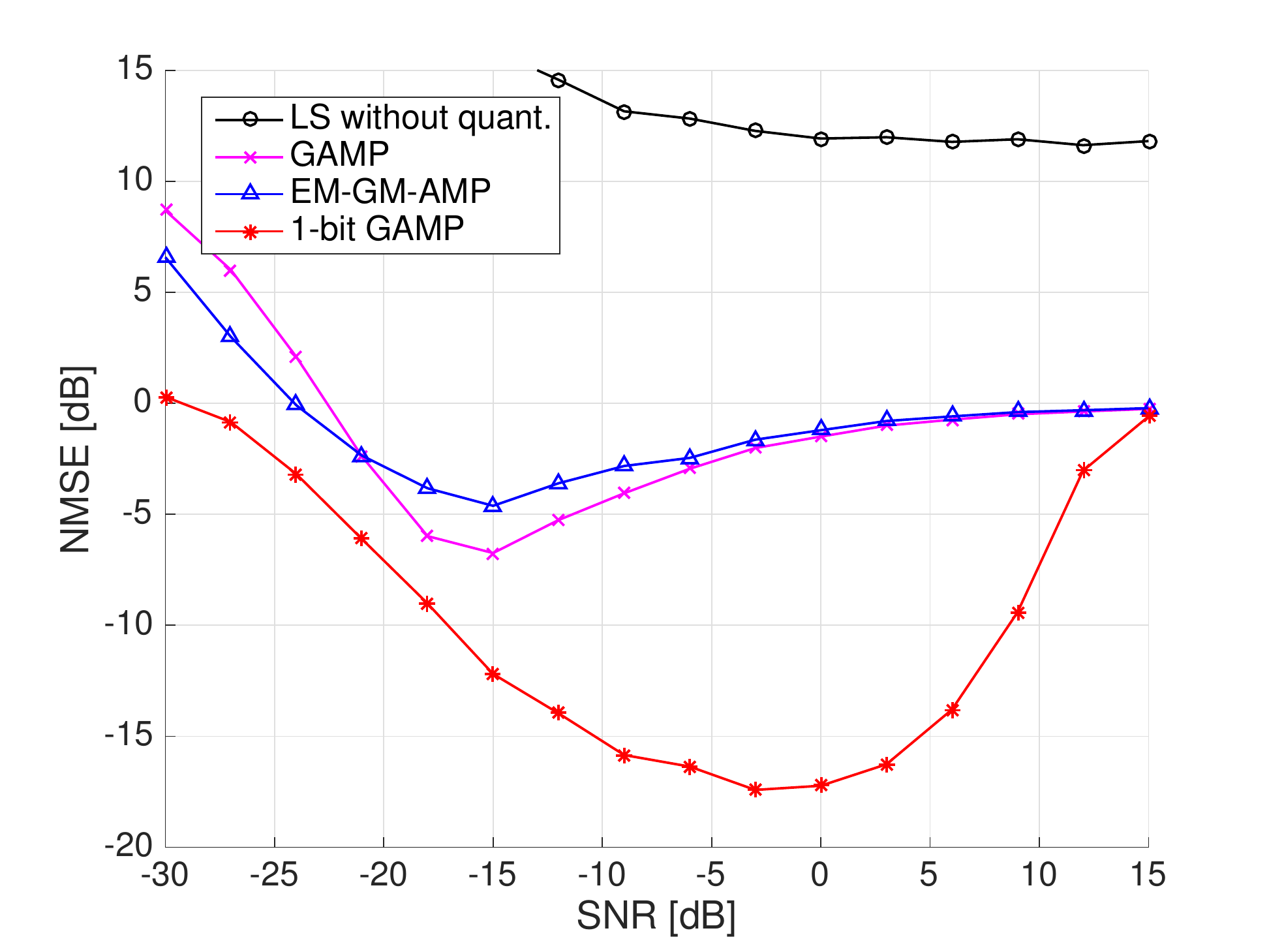}}
	  % \label{fig:unq_nmse_comparison}
	%  \vspace{2.0cm}
	  \centerline{(a)}\medskip
	  
	\end{minipage}
	\begin{minipage}[b]{1.0\linewidth}
	  \centering
	  \centerline{\includegraphics[width=7.4cm]{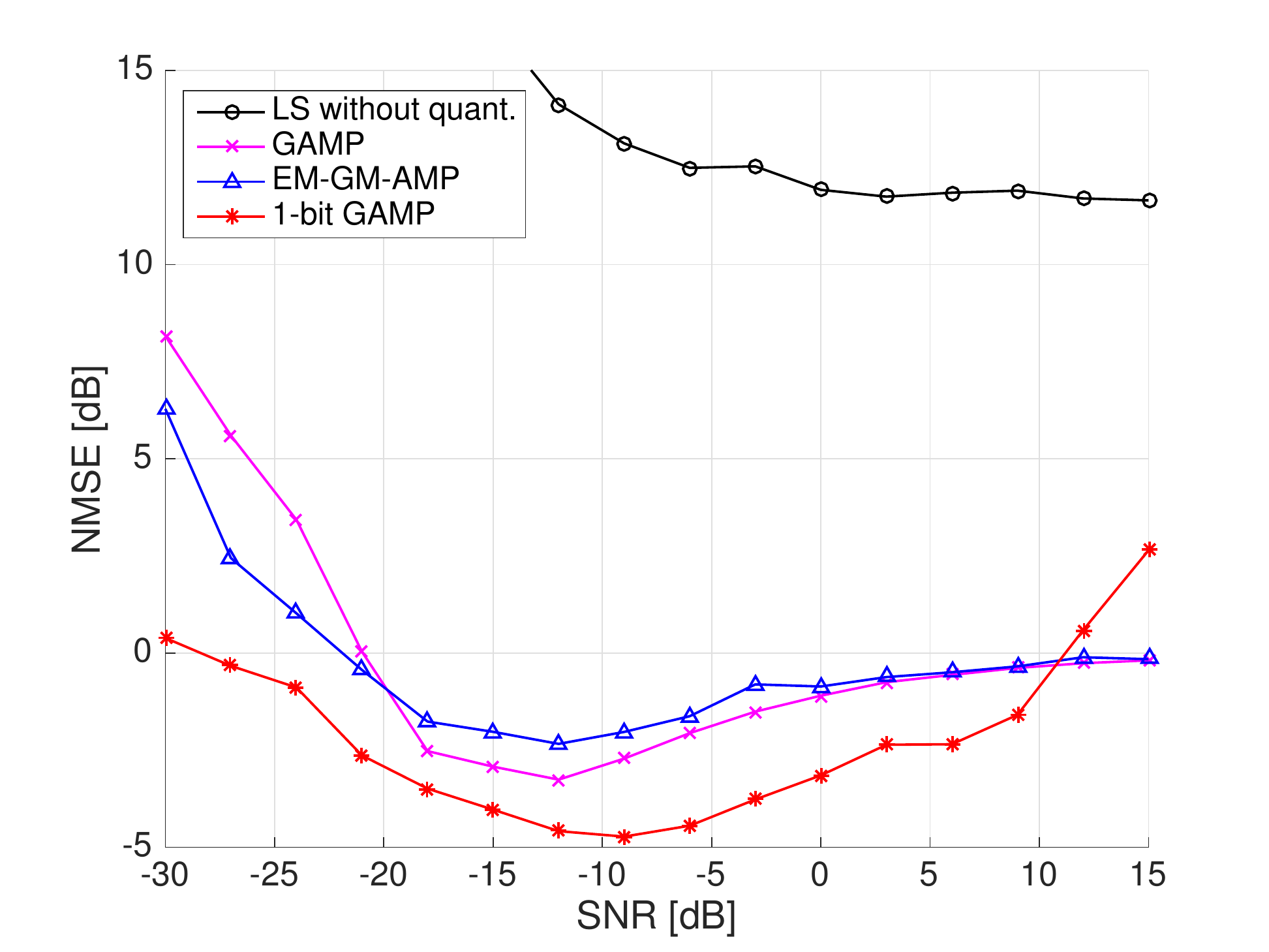}}
	%  \vspace{2.0cm}
	  \centerline{(b)}\medskip
	  % \label{fig:quant_nmse_comparison}
	\end{minipage}
	\vspace{-.3in}
	\caption{NMSE of four channel estimation algorithms: LS, GAMP, EM-GM-AMP and one-bit GAMP (proposed). The leakage effect is considered in (b) and degrades channel estimation performance.}
	\label{fig:nmse_comparison}
\end{figure}

\section{Numerical Results} % (fold)
\label{sec:numerical_results}
In this section, we evaluate channel estimation performance of the modified one-bit GAMP and compare it with other algorithms that include LS estimation, GAMP, and EM-GM-AMP \cite{projectcode}. The normalized mean squared error (NMSE) is used as a performance metric:
\begin{align*}
	%\text{NMSE} = \mathbb{E}\left[ \frac{\left\lVert \mathbf{H} - \hat{\mathbf{H}}\right\lVert_{F}^2}{\left\lVert \mathbf{H}\right\lVert_{F}^2} \right].
	\text{NMSE} = \mathbb{E}\left[ {\lVert \mathbf{H} - \hat{\mathbf{H}}\rVert_{F}^2}/{\left\lVert \mathbf{H}\right\lVert_{F}^2} \right].
\end{align*}
For simulation, system parameters used in this section are as follows unless otherwise stated: $N_t=64$, $N_r=16$, $L_t=4$, $L_r=4$, $N_p=2$, and $M=64$. Columns of a Hadamard matrix are used for training symbol vectors. 
%To highlight the performance of one-bit GAMP, NMSE of other algorithms are plotted in Fig.~\ref{fig:nmse_comparison}. 
%The algorithms employed for the comparison purposes are least-square (LS) estimation, GAMP \cite{Rangan2011}, and EM-GM-AMP \cite{Vila2013}. 

Fig.~\ref{fig:nmse_comparison} shows NMSE of four channel estimation algorithms. For comparison purposes, the LS estimator is without signal quantization while others use one-bit ADCs. Simulated channels for Fig.~\ref{fig:nmse_comparison}(a) are intentionally constructed to avoid the leakage effect whereas those for Fig.~\ref{fig:nmse_comparison}(b) do not have such constraint. As seen in both subfigures, the LS estimator without quantization achieves far worse performance than do GAMP variants with one-bit ADCs. One-bit GAMP yields the lowest estimation errors among the considered algorithms in both subfigures. Comparing two subfigures, we can see that all algorithms suffer performance degradation due to the leakage effect.
%For all variants of GAMP, NMSE is not a monotonic function of SNR as seen in the figure, and this phenomenon is called the stochastic resonance \cite{mcdonnell2008stochastic,Mo2014}.  All algorithms except LS suffer from performance degradation due to the leakage effect. In both cases, one-bit GAMP achieves better performance than the others, and this performance gap is prominent without the leakage effect. 
%Note that even though GAMP and EM-BG-AMP give similar performance, EM-BG-AMP has a merit that estimates required parameters at the cost of extra computation. 
Figures henceforth are with one-bit GAMP and channels that experience the leakage effect.  

\begin{figure}[t]
	\centering
	\centerline{\includegraphics[width=7.4cm]{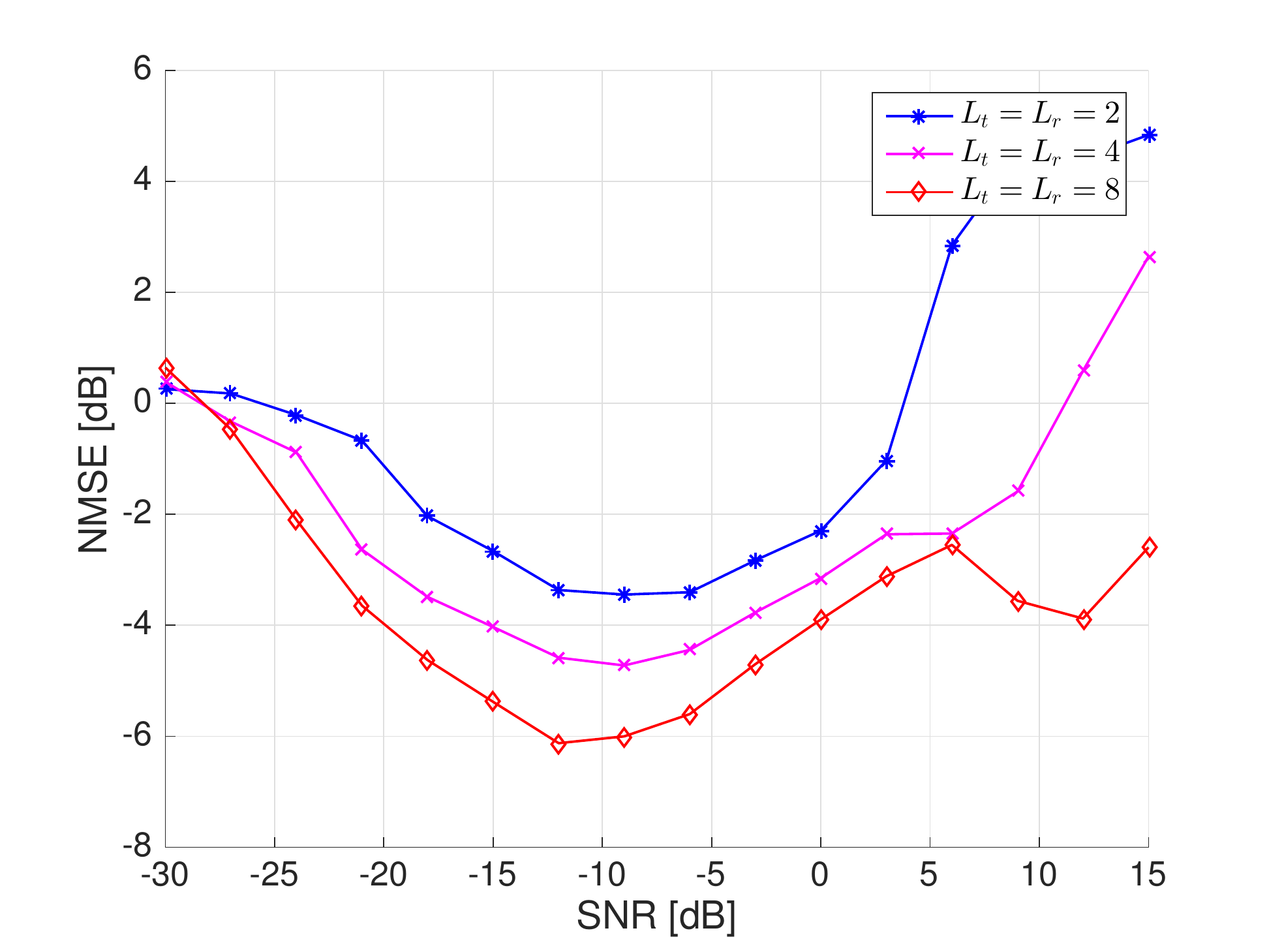}}
	\caption{NMSE as a function of RF chains. More RF chains improve channel estimation performance.}
	\label{fig:quant_nmse_vs_rfchains}
\end{figure}

The number of RF chains affects the channel estimation performance as shown in Fig.~\ref{fig:quant_nmse_vs_rfchains}. Two, four and eight pairs of RF chains are plotted. More RF chains improve the performance across the considered SNR range. From a compressed sensing perspective, this is because more RF chains allow longer measurement vector.

\begin{figure}[t]
	\centering
	\centerline{\includegraphics[width=7.4cm]{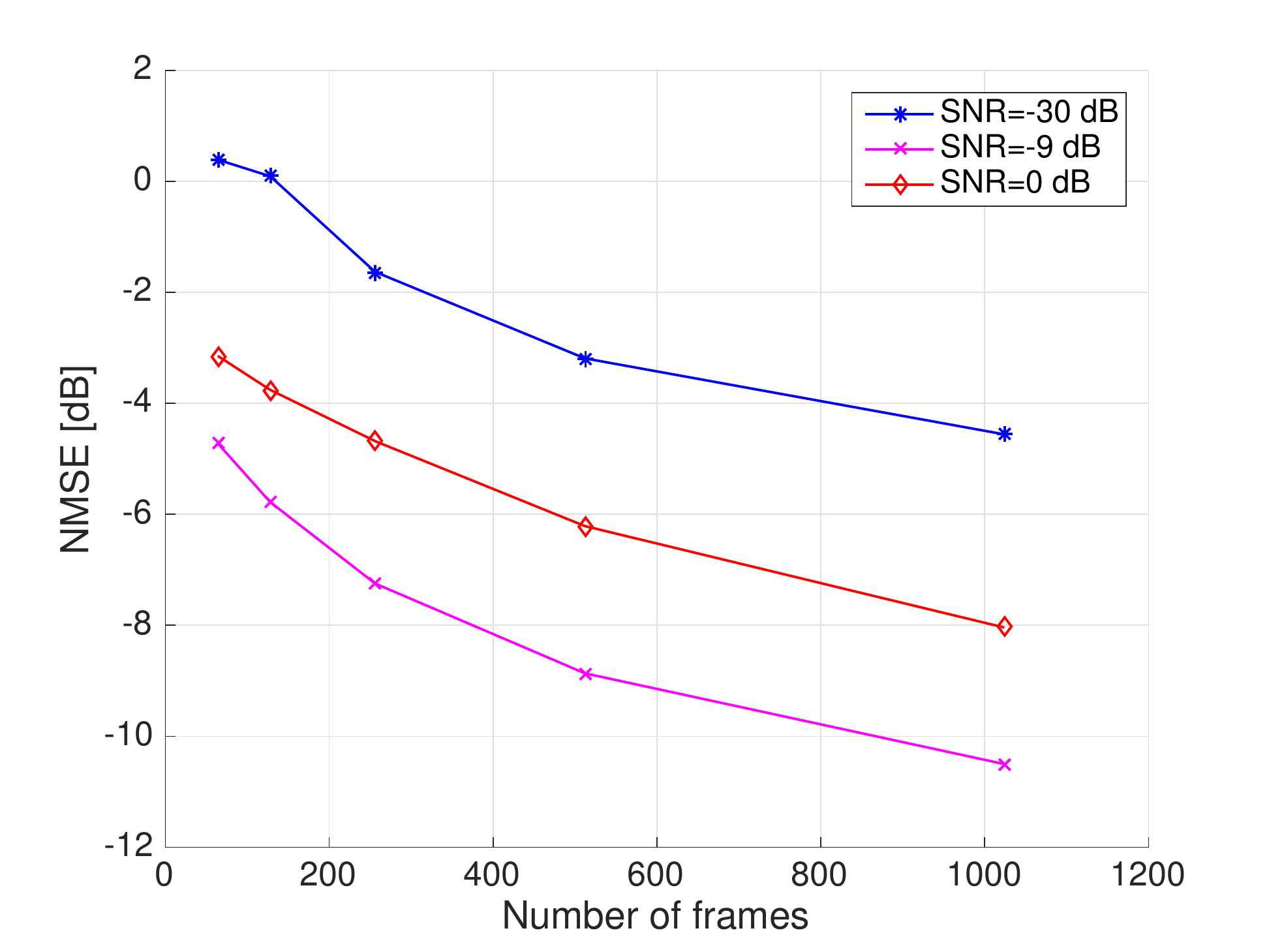}}
	\caption{NMSE as a function of frames. For any SNR, estimation error decreases with increasing number of frames.}
	\label{fig:quant_nmse_vs_frames}
\end{figure}

Fig.~\ref{fig:quant_nmse_vs_frames} shows effects of the number of frames on channel estimation performance in various SNR regimes. As shown in Fig.~\ref{fig:nmse_comparison}(b), the minimum NMSE can be obtained around an SNR of -9 dB, %and this SNR value always fulfills the lowest NMSE across different number of frames. Thus 
which corresponds to the fact that the curve for -9 dB SNR in Fig.~\ref{fig:quant_nmse_vs_frames} has lower NMSE than the others. Regardless of SNR values, estimation error decreases as more frames are used for estimation. It is expected since coherence of the measurement matrix declines with the increasing number of frames for any SNR. 
% \begin{figure}[t]
% 	\centering
% 	\centerline{\includegraphics[width=7.4cm]{plots/quant_sparsity_vs_nmse}}
% 	\caption{NMSE vs. Channel sparsity}
% 	\label{fig:qant_nmse_vs_sparsity}
% \end{figure}

% and it leads to finer granularity in AoA and AoD estimation. 
%This fact also can be supported by plotting NMSE with the aligned steering vector case where three plots cross each other across the SNR range. The plot is not represented in this paper due to the space limit.

% The last factor we consider is the sparsity of the mmWave channel. It can be seen from the curves in Fig.~\ref{fig:qant_nmse_vs_sparsity} that, regardless of SNR value, NMSE increases with sparsity. 

\section{Conclusion} % (fold)
\label{sec:conclusion}
In this paper, we proposed a channel estimation algorithm for a mmWave communication system with one-bit quantizers and hybrid beamforming based on GAMP. This one-bit GAMP method is specifically designed for measurements taken with one-bit ADCs, and we modified it to take into account the thermal noise. Simulation results showed that GAMP variants with one-bit ADCs achieve better performance than LS without quantization, and that the proposed algorithm yields the lowest channel estimation error among the GAMP variants. Results also showed that the channel estimation performance can be enhanced by exploiting more frames and RF chains.

%We provided simulation results that show the proposed algorithm achieves superior performance to other widely used traditional and sparse recovery channel estimation algorithms such as LS, GAMP and EM-GM-GAMP using NMSE as a performance metric. Results also showed that the channel estimation performance can be enhanced with %lower channel sparsity and 
%more frames and RF chains.

% section conclusion (end)
% \begin{figure}[htb]
% 	\centering
% 	\centerline{\includegraphics[width=8.5cm]{plots/unq_frame_vs_nmse}}
% 	\caption{NMSE vs. Number of frames}
% 	% \label{fig:unq_nmse_comparison}
% \end{figure}

% \begin{figure}[htb]
% 	\centering
% 	\centerline{\includegraphics[width=8.5cm]{plots/unq_rfChains_vs_nmse}}
% 	\caption{NMSE vs. Number of RF chains}
% 	% \label{fig:unq_nmse_comparison}
% \end{figure}

% \begin{figure}[htb]
% 	\centering
% 	\centerline{\includegraphics[width=8.5cm]{plots/unq_sparsity_vs_nmse}}
% 	\caption{NMSE vs. Channel sparsity}
% 	% \label{fig:unq_nmse_comparison}
% \end{figure}
% section numerical_results (end)

% \cite{C2}
\vfill\pagebreak

\bibliographystyle{IEEEbib}
\bibliography{strings,refs}

\end{document}